\documentclass[12pt]{article}
\usepackage{amstex}

\def\Htilde{\tilde{H}}
\def\Ytilde{\tilde{Y}}

\def\ep{\epsilon}

\def\ie{{\it i.e.}}

\begin{document}
\vspace*{-.6in}
\thispagestyle{empty}
\begin{flushright}
CALT-68-2124\\
hep-th/9706197
\end{flushright}
\baselineskip = 20pt

\vspace{.5in}
{\Large
\begin{center}
The M Theory Five-Brane\footnote{Lecture presented at the APCPT--ICTP Joint
International Conference '97 (AIJIC97), May 26--30, 1997 in Seoul Korea.}
\end{center}}

\begin{center}
John H. Schwarz\\
\emph{California Institute of Technology, Pasadena, CA  91125, USA}
\end{center}
\vspace{0.1in}

\begin{center}
\textbf{Abstract}
\end{center}
\begin{quotation}
\noindent  BPS saturated p-branes play an important role in recent
progress in understanding superstring theory and M theory. One approach to understanding
the dynamics of p-branes is to formulate an effective (p+1)-dimensional
world-volume theory. The construction of such
brane actions involves a number of interesting issues.
One such issue is how to formulate
the action for theories that contain chiral bosons. 
The two main examples, which are the M theory five-brane
and the heterotic string, are described in this lecture. Also, double dimensional reduction of the
M theory five-brane on K3 is shown to give the heterotic string.
\end{quotation}
\bigskip

\section{Introduction}

In the first superstring revolution (1984--85) we learned that there are five consistent 
superstring theories, each of which requires ten-dimensional space-time. Each of these
theories, is approximated at low energy by an effective 10d supergravity theory, yet
it has a consistent perturbation expansion, free from ultraviolet divergences,
based on the appropriate fundamental string. The five theories are:

\noindent{\it Type I}: this theory has N=1 supersymmetry (a Majorana--Weyl supercharge)
and SO(32) gauge symmetry; type I strings are unoriented and can be open or closed.

\noindent The remaining four theories all are based on fundamental strings that are oriented and closed.

\noindent{\it Type IIA}: this theory has N=2A supersymmetry (a pair of Majorana--Weyl
supercharges of opposite chirality) and no gauge symmetry.

\noindent{\it Type IIB}: this theory has N=2B supersymmetry (a pair of Majorana--Weyl
supercharges with the same chirality) and no gauge symmetry.

\noindent{\it HO}: the heterotic string theory with N=1 supersymmetry and SO(32)
gauge symmetry.

\noindent{\it HE}: the heterotic string theory with N=1 supersymmetry and $E_8 \times E_8$
gauge symmetry.

In the second superstring revolution (1994 -- ?)  we have learned that all five
superstring theories are actually different limiting cases of a single underlying 
theory. (For a review, see ref.~\cite{jhsreview}.)
In other words, they are nonperturbatively equivalent. The way this
works is that they are related by various dualities (discussed below) such that
fundamental states in one description can appear as solitons of a dual description.
(For example, the HO string is a D-string of the Type I theory.) Moreover, a 10th spatial
dimension arises nonperturbatively in the IIA and HE theories, so that at strong coupling these
theories are actually 11-dimensional. Thus in the limit one obtains a vacuum with
11d super-Poincar\'e invariance, a configuration that is highly nonperturbative
from the string viewpoint. The quantum theory with this vacuum is called M theory. It is 
still rather mysterious, though we know some facts about it, for example that it is
approximated at low energies by 11d supergravity. A clever proposal for a fundamental
description -- tentatively called Matrix Theory -- was made 
within the past year~\cite{banks}. It
looks quite promising, though it seems still to be incomplete when too many dimensions
are compactified. In any case, it is currently under intense scrutiny, and progress in
understanding is occurring rapidly. 

\medskip

\noindent{\it Dualities}

The dualities that relate the various string theories and M theory are designated by
the letters S, T, and U. Two theories, call them A and B, are said to be S dual if theory A 
evaluated at strong coupling is equivalent to theory B at weak coupling and 
vice versa.  This means that one coupling constant is the reciprocal of the other.
In string theory the coupling constant is given by the vev of a scalar field, called the dilaton,
by $\lambda = <e^{\phi}>$. Thus if A and B are S-dual, their dilaton fields are
related by $\phi_A = - \phi_B$.

To understand T duality consider two theories, again called A and B, that are defined
on manifolds $M \times K_A$ and $M \times K_B$, respectively, where $M$ represents
space-time and $K_A$ and $K_B$ are compact internal spaces. Then, A and B are T 
dual if the two theories are physically equivalent with a correspondence between
$K_A$ and $K_B$ such that the volume of one is inversely proportional to the
volume of the other. The volumes are also given by the vevs of scalar fields by $V = <e^{\psi}>$,
so the volume moduli are related by $\psi_A = - \psi_B$, which is quite analogous
to the relation between dilatons of an S-dual pair.

The notion  of U duality combines aspects of S and T. Theories A and B are U dual if 
theory A with large (or small) compact space $K_A$ is equivalent to theory B at strong
(or weak) coupling. This means that  $\psi_A = \pm \phi_B$.

In the special case where A = B, the duality is a symmetry (a discrete gauge symmetry, in fact).
For example, type IIB superstring theory has an SL(2, Z) group of duality symmetries,
one of whose elements is an S duality transformation.

\medskip

\noindent{\it BPS States}

Another ingredient that has played an important role in recent progress is the
identification of BPS states. This is the technical tool that allows us to extract
nonperturbative information about theories that originally were only known perturbatively.
The basic idea is that in systems with sufficient supersymmetry, there are conserved
charges that appear in the supersymmetry algebra (in addition to the supercharges
and the momentum). Particles that carry these charges have a mass that is
bounded below as a consequence of the algebra. (Suitably normalized, one has
$M \geq |Q|$, where $Q$ is the charge.) When the bound is saturated, the state is called a
BPS state, and one can prove that it belongs to a ``short" representation of the
supersymmetry algebra. (This generalizes the well-known fact that a photon in 4d
has two polarizations rather than three.) The utility of short representations is that
the relation between charge and mass must be maintained as the strength of the coupling is
increased or other moduli are varied, so long as there is no phase transition.
This story has a straightforward generalization to extended p-dimensional objects,
called p-branes. Their tensions (mass per unit volume) satisfy analogous inequalities
and when the tension equals its minimum allowed value one again has a short representation.

\medskip

\noindent{\it p-branes}

The effective supergravities in question contain various antisymmetric tensor
gauge fields.  These can be represented as differential forms
\begin{equation}
A_n = A_{\mu_{1}\mu_{2} \ldots \mu_{n}} dx^{\mu_{1}}~\wedge dx^{\mu_{2}}
\wedge \ldots \wedge dx^{\mu_{n}} .
\end{equation}
In this notation, a gauge transformation is given by $\delta A_n = d \Lambda_{n
- 1}$, and the gauge-invariant field strength is $F_{n + 1} = dA_n$.  When
interactions are included, these formulas are sometimes modified.  The origin
of $p$-branes can be understood by considering an action that (schematically)
has the structure~\cite{horowitz91}
\begin{equation}
S \sim \int d^D x \sqrt{-g} \{R + (\partial \phi)^2 + e^{-a\phi} F_{n+1}^2
+ \ldots\}.
\end{equation}
Here, $\phi$ represents a dilaton field, $R$ is the scalar curvature, and $a$
is a numerical constant whose value depends on the particular theory.  The dots
include all the additional terms required to make the theory locally
supersymmetric.  In this case it is meaningful to seek BPS
$p$-brane solutions, and it turns out that solutions exist for $p = n - 1$ and
$p = D - n - 3$.  By a straightforward generalization of the nomenclature of
Maxwell theory, it is natural to call these ``electric'' and ``magnetic,''
respectively.  The electric $p$-brane, with $p = n - 1$, has an $n$-dimensional
world-volume.  The fact that it is a source for ``electric'' charge is
exhibited by the coupling
\begin{equation}
\int A_{\mu_{1} \ldots \mu_{n}} {\partial x^{\mu_{1}}\over \partial \sigma^1}
\ldots {\partial x^{\mu_{n}}\over\partial\sigma^n} d^n \sigma,
\end{equation}
which generalizes the familiar $j \cdot A$ coupling of Maxwell theory.

A $p$-brane in $D$ dimensions (let's assume it is an infinite hyperplane, for
simplicity) can be encircled by a $(D-p-2)$-dimensional sphere $S^{D-p-2}$.
Thus, the ``electric charge'' of the $p$-brane is given by a straightforward
generalization of Gauss's law for point charges
\begin{equation}
Q_E \sim \int_{S^{D - p - 2}} * F,
\end{equation}
where $*F$ is the Hodge dual of $F$.  In these lectures, we will not need to
commit ourselves to specific normalization conventions.  Similarly, a dual $(D
- p - 4)$-brane has ``magnetic charge''
\begin{equation}
Q_M \sim \int_{S^{p + 2}} F.
\end{equation}
Note that the charge associated with a $p$-brane has dimension $(length)^{D/2 -
2 - p}$.  
This is dimensionless when $p = (D-4)/2$ -- \ie, for point particles in
4d, strings in 6d, membranes in 8d, etc.  In these cases the
electric and magnetic branes have the same dimensionality and it is possible to
have dyonic $p$-branes.

The charges of $p$-branes can also be described by generalizations of Coulomb's
law.  So, for an electric $p$-brane, as $r\rightarrow\infty$
\begin{equation}
A \sim  {Q_E \over r^{D - p - 3}}\omega_{p + 1},
\end{equation}
where $r$ is the transverse distance from the brane and $\omega_{p+1}$ is the
volume form for the $p$-brane world-volume.  Similarly, for the dual magnetic
($D -p-4$)-brane, as $r\rightarrow\infty$
\begin{equation}
F \sim  {Q_M \over r^{p + 2}}\Omega_{p + 2},
\end{equation}
where $\Omega_{p+2}$ is the volume form on a sphere $S^{p + 2}$ surrounding the
brane.  In this case it is convenient to describe the magnetic field, rather
than the potential, in order to avoid introducing generalizations of Dirac
strings.  Of course, the distinction between electric and magnetic branes is
not so great, since it is often possible to make a duality transformation that
replaces $A$ by a dual potential $\tilde{A}$ whose field strength $d\tilde{A}$
is the dual of $F = dA$.  From the point of view of $\tilde{A}$, the original
electric brane is magnetic and vice versa.  Another significant fact,\cite{nepomechie85} 
noted more  than ten years ago, is that the Dirac quantization condition has a
straightforward generalization to the charges carried by a dual pair of
$p$-branes: $Q_E Q_M \in 2\pi {\bf Z}$.  This assumes appropriate normalization
conventions, of course.

The crudest first approximation to classical $p$-brane dynamics is given by a
straightforward generalization of the Nambu area formula
for the string world-sheet action.  This gives an action
proportional to the $(p + 1)$-dimensional volume induced by
embedding  the world volume into the $D$-dimensional target space:
\begin{equation}
S_{eff} = T_P \int \sqrt{{\rm det}\ G_{\alpha\beta}}\, d^{p + 1} \sigma ,
\end{equation}
where
\begin{equation}
G_{\alpha\beta} = \eta_{\mu\nu} \partial_\alpha X^\mu \partial_\beta
X^\nu,  \alpha,\beta = 0, 1, \ldots, p,
\end{equation}
and $\eta$ is the metric (Minkowski, for example) of the target space.  Just
as for strings, this formula is invariant under reparametrizations of the world
volume.  Also, it defines the $p$-brane tension $T_p$ -- the universal mass per
unit volume of the $p$-brane.  Note that $T_p \sim (mass)^{p +1}$.

A significant class of p-branes that has played a major role in recent developments
are called D-branes (or Dp-branes)~\cite{polchinski}. 
(These only occur for the type I and type II theories.)
The D stands for ``Dirichlet'', because these
p-branes are defined in terms of open strings whose endpoint boundary conditions
are Neumann in certain directions and Dirichlet in others. The Dirichlet boundary
conditions force the brane to end on a hypersurface, which turns out to be a dynamical
object.  A nifty thing about D-branes is that much of their dynamics can be
understood in terms of the dynamics of the open strings that define them.
D-branes are not the subject of this lecture, so let me just list a few of their salient  
properties: 1) Their tensions (measured in the string metric) are proportional to $1/\lambda$,
where $\lambda$ is the string couping constant. This behavior is intermediate between
that of a fundamental string, whose tension is $\lambda$ independent, and all other
solitons, whose tensions are proportional to $1/\lambda^2$.  2) The conserved charge
carried by a D-brane arises from the Ramond--Ramond sector of the theory. This
means that they can be viewed as bispinors. 3) The D-brane world volume theory
contains a U(1) gauge field.  When a string ends on a D-brane there is an
``electric" charge on its end, which creates a Coulomb-like field in the brane.
The dynamics of a set of $n$ parallel identical D-branes can be described 
by a U(n) gauge theory. The off-diagonal fields of the matrix arise from the ground
states of open strings connecting pairs of D-branes. These are massive, of course, when the
branes are not coincident.

You might wonder why there is so much emphasis on p-branes of late. The fact
is that they have been invoked in a number of quite different settings. One viewpoint
is that just as perturbative string theories were based on strings,
perturbative expansions might be based on objects of other dimensions.
In the case of fundamental point particles, this is just ordinary quantum field
theory. The more radical suggestion, that M theory could be defined in terms
of fundamental supermembranes (2-branes) was popular for a while, but does not
seem promising at this time. Another viewpoint, which seems to make more sense,
is that when one views the theory nonperturbatively, all BPS saturated objects
have a similar algebraic and dynamical status (though each has its own peculiarities).
This led Townsend to introduce the notion of ``p-brane democracy"~\cite{townsend2}.

The most important uses of p-branes (other than fundamental strings as the basis
of perturbation expansions) are the following:  1) Vacuum 
configurations that are nonperturbative from
a string theory viewpoint can be defined by introducing p-branes that completely
fill the noncompact space-time dimensions.
Interesting examples of this are the so-called F theory vacua~\cite{vafa}. 
They are nonperturbative
type IIB vacua which contain 24 7-branes in a consistent manner. 
2) A variation on the preceding theme is to formulate a consistent brane
configuration for the purpose of studying the gauge field theory that lives on the
branes~\cite{hanany}. This is a rapidly developing subject that is leading to dramatic
progress in understanding nonperturbative properties of supersymmetric gauge theories
in a very geometrical sort of way. Examples will be described in Shapere's contribution
to this conference. 3) Branes can be introduced into specific string vacua as 
``probes"~\cite{banks2}.
When they approach other branes or singularities in the space-time geometry,
interesting dynamics is induced in the world volume of the probe. This can be
a very interesting theory in its own right, or it can tell us interesting things about
the space it is probing. Gauge symmetries of the space-time theory appear as global 
symmetries of the theory on the probe. 4) Finite volume branes can wrap around
around cycles of compact dimensions. This describes a class of excitations of the theory.
An especially interesting class of examples is wrapped D-branes that correspond to
black holes when a coupling constant is continues from weak to strong coupling.
This can be done in a controlled way for extremal/BPS  cases. In a large class of
examples of this type one has been able to count microscopic states, defined as 
excitation of the D branes, and to show that the counting agrees with the
classical Bekenstein--Hawking entropy formula. This program has been carried
quite far with studies of the Hawking radiation, deviations from a thermal
distribution, extensions to nonextremal black holes, etc. -- all with impressive
success.

\medskip

\noindent{\it The Branes of M Theory}

The massless fields of M theory are just those of 11-dimensional supergravity:
the metric tensor, the gravitino, and a three form potential $A_3$. By the
reasoning explained above, there are two kinds of BPS p-branes that can couple
to the three-form potential. The one that couples electically is the M2-brane,
originally called the supermembrane. Its world volume theory was constructed
ten years ago~\cite{bergshoeff}. The brane that couples
magnetically to $A_3$ has five spatial dimensions and is called the M5-brane.
Its world volume theory, which was constructed very recently will be
described below. The description of the fermionic degrees of the M5-brane
involves a number of technical issues that I do not have time to get into here.
So I will only describe the bosonic truncation of the M5-brane action. I should
emphasize, however,  that the complete action with global 11d supersymmetry
and local kappa symmetry on the world sheet has been constructed.

\section{The Bosonic Part of the Five-Brane Action}

\noindent{\it The World-Volume Field Content}

The presence of an M5-brane breaks half of the supersymmetry of the
background 11d supergeometry. The Majorana supercharge of eleven dimensions
has 32 components. It decomposes on a six-dimensional subspace into two
positive chirality and two negative chirality spinors. One can argue in this case
that both of the unbroken 6d supersymmetries have the same chirality and therefore
the 6d world volume theory of the M5-brane has (2,0) supersymmetry.
Corresponding to the broken supersymmetries one has masssless Goldstone
fermions in the world volume theory. They give 8 physical degrees of freedom.
There is only one (2,0) supermultiplet with this content and it is called the
tensor multiplet. Its bosonic fields are a two-form potential with a self-dual
field strength and five scalar fields. The scalar fields can be interpreted as
the Goldstone bosons associated with the five translation symmetries, in directions
normal to the five-brane, broken by the presence of the five-brane.  In a 
covariant description they are represented by the 11  coordinates
$X^M$ that describe the embedding of the brane into the spacetime.
The fact that the longitudinal components are nondynamical is built in by
constructing the world volume theory to have 6d general coordinate invariance.
Thus they could be eliminated by passing to a physical gauge, though we will not
do that here. Supersymmetry requires that the number of propagating bosons
on the M5-brane should equal the number of propagating fermions, which is
eight. Thus the self-dual gauge field should have three propagating modes.
That this is the case is most easily understood by noting that it belongs to the
(3,1) representation of the $SO(4) = SU(2) \times SU(2)$ little group in
six dimensions. A parity transformation interchanges the two $SU(2)$'s, and
so we see that it is a chiral boson.

Ref. \cite{jhs} analyzed the problem of coupling a 6d self-dual tensor gauge field to a
metric field so as to achieve general coordinate invariance. 
It presented a formulation in which one direction is treated differently from
the other five. At the time that work was done,
the author knew of no straightforward way to make the general
covariance manifest. However, shortly thereafter a paper appeared~\cite{pasti1} that 
presents equivalent results using a manifestly covariant formulation~\cite{pasti2},
which we refer to as the PST formulation. In the following both approaches and their
relationship are described. These results have been generalized to
supersymmetric actions with local kappa symmetry~\cite{bandos,aganagic,howe}, 
but here we will only consider the bosonic theories.

\medskip

\noindent{\it The Noncovariant Formulation}

Let us denote the 6d (world volume) coordinates by 
$\sigma^{\hat\mu} = (\sigma^\mu, \sigma^5)$,
where $\mu = 0,1,2,3,4$. The $\sigma^5$ direction is singled out as the one that
will be treated differently from the other five.\footnote{This is a
space-like direction, but one could also choose a time-like  one.}  The 6d metric
$G_{\hat\mu\hat\nu}$ contains 5d pieces $G_{\mu\nu}, G_{\mu 5}$, and $G_{55}$.
All formulas will be written with manifest 5d general coordinate invariance.
As in refs.~\cite{perry,jhs}, we represent the self-dual tensor gauge field by a
$5\times 5$ antisymmetric tensor $B_{\mu\nu}$, and its 5d curl by
$H_{\mu\nu\rho} = 3 \partial_{[\mu} B_{\nu\rho]}$. A useful quantity is the dual 
\begin{equation}
\tilde{H}^{\mu\nu} = {1\over 6} \epsilon^{\mu\nu\rho\lambda\sigma}
H_{\rho\lambda\sigma}.
\end{equation}

It was shown in ref.~\cite{jhs} that a class of generally covariant
bosonic theories can be represented in the form
$L = L_1 + L_2 + L_3$, where
\begin{eqnarray}
L_1 &=& -{1\over 2}\sqrt{-G} f(z_1,z_2), \nonumber \\
L_2 &=& -{1\over 4} \tilde{H}^{\mu\nu} \partial_5 B_{\mu\nu}, \\
L_3 &=&  {1\over 8}
\epsilon_{\mu\nu\rho\lambda\sigma} {G^{5\rho}\over G^{55}} \tilde{H}^{\mu\nu}
\tilde{H}^{\lambda\sigma}.\nonumber 
\end{eqnarray}
The notation is as follows:  $G$ is the 6d determinant $(G =
{\rm det}\, G_{\hat\mu\hat\nu})$ and
$G_5$ is the 5d determinant $(G_5 =
{\rm det}\, G_{\mu\nu})$, while $G^{55}$ and $G^{5\rho}$ are components of the inverse
6d metric $G^{\hat\mu \hat\nu}$.  The $\epsilon$ symbols are purely numerical with $\epsilon^{01234} = 1$ and
$\epsilon^{\mu\nu\rho\lambda\sigma} = - \epsilon_{\mu\nu\rho\lambda\sigma}$.  A
useful relation is $G_5 = G G^{55}$.
The $z$ variables are defined to be
\begin{eqnarray}
z_1 &=& {{\rm tr} (G\tilde{H} G\tilde{H})\over 2( -G_5)}\nonumber \\
z_2 &=& {{\rm tr} (G\tilde{H} G\tilde{H} G\tilde{H} G\tilde{H})\over 4 (-G_5)^2}.
\label{zdefs}
\end{eqnarray}
The trace only involves 5d indices:
\begin{equation}
{\rm tr} (G\tilde{H} G\tilde{H}) = G_{\mu\nu} \tilde{H}^{\nu\rho} G_{\rho\lambda}
\tilde{H}^{\lambda\mu}.
\end{equation}
The quantities $z_1$ and $z_2$
are scalars under 5d general coordinate
transformations.  

Infinitesimal parameters of general coordinate transformations are denoted
$\xi^{\hat\mu} = (\xi^\mu, \xi)$.  Since 5d general coordinate invariance is
manifest, we focus on the $\xi$ transformations only.  The metric transforms in
the standard way
\begin{equation}
\delta_\xi G_{\hat\mu \hat\nu} = \xi \partial_5 G_{\hat\mu \hat\nu} +
\partial_{\hat\mu} \xi G_{5\hat\nu} + \partial_{\hat\nu} \xi G_{\hat\mu 5}.
\label{Gvar}
\end{equation}
The variation of $B_{\mu\nu}$ is given by a more complicated rule, whose origin is
explained in ref.~\cite{jhs}:
\begin{equation}
\delta_\xi B_{\mu\nu} = \xi K_{\mu\nu}, \label{Bvar}
\end{equation}
where
\begin{equation}
K_{\mu\nu} = 2{\partial (L_1 + L_3) \over\partial \tilde{H}^{\mu\nu}} =
K_{\mu\nu}^{(1)} f_1+ K_{\mu\nu}^{(2)} f_2+ K_{\mu\nu}^{(\epsilon)}
\label{Kform1}
\end{equation}
with
\begin{eqnarray}
K_{\mu\nu}^{(1)} &=& {\sqrt{-G} \over (-G_5)}{(G\tilde{H} G)_{\mu\nu}}
\nonumber \\
K_{\mu\nu}^{(2)} &=& {\sqrt{-G} \over (-G_5)^2}{(G\tilde{H} G\tilde{H} G\tilde{H}
G)_{\mu\nu}}  \label{Kform2}\\
K_{\mu\nu}^{(\epsilon)} &=& \epsilon_{\mu\nu\rho\lambda\sigma}
{G^{5\rho}\over 2 G^{55}} \tilde{H}^{\lambda\sigma}, \nonumber
\end{eqnarray}
and we have defined
\begin{equation}
f_i = {\partial f\over\partial z_i} , \quad i = 1,2.
\end{equation}

Assembling the results given above, ref.~\cite{jhs} showed that
the required general coordinate transformation symmetry is
achieved if, and only if, the function $f$ satisfies the nonlinear partial
differential equation~\cite{gibbons}
\begin{equation} \label{nlpde}
f_1^2 + z_1 f_1 f_2 + \big({1\over 2} z_1^2 - z_2\big) f_2^2 = 1.
\end{equation}
As discussed in~\cite{perry} and described in an appendix,
this equation has many solutions, but the one of relevance to the
M theory five-brane is 
\begin{equation}
f = 2 \sqrt{1 + z_1 + {1\over 2} z_1^2 - z_2}.
\end{equation}
For this choice $L_1$
can reexpressed in the Born--Infeld form
\begin{equation}
L _1 = - \sqrt{- {\rm det} \Big(G_{\hat\mu \hat\nu} + i G_{\hat\mu\rho} G_{\hat\nu
\lambda} \tilde{H}^{\rho\lambda} / \sqrt{-G_5}\Big)} . \label{bosonicL1}
\end{equation}
This expression is real, despite the factor of $i$, because it is an even function of
$\tilde H$.

\medskip

\noindent{\it The PST Formulation}

In ref.~\cite{pasti1} (using techniques developed
in ref.~\cite{pasti2}) equivalent results are
described in a manifestly covariant way.  To do this, the field $B_{\mu\nu}$ is
extended to $B_{\hat\mu \hat\nu}$ with field strength $H_{\hat\mu \hat\nu
\hat\rho}$.  In addition, an auxiliary scalar field $a$ is
introduced.  The PST formulation has new gauge symmetries (described below)
that allow one to choose the gauge $B_{\mu 5} = 0,$  $a = \sigma^5$
(and hence $\partial_{\hat\mu}a =
\delta_{\hat\mu}^5$).  In this gauge, the covariant PST formulas reduce to the
ones given above.

Equation (\ref{bosonicL1}) 
expresses $L_1$ in terms of the determinant of the $6 \times 6$ matrix
\begin{equation}
M_{\hat\mu\hat\nu} = G_{\hat\mu\hat\nu} + i {G_{\hat\mu \rho} G_{\hat\nu
\lambda}\over \sqrt{- GG^{55}}} \tilde{H}^{\rho\lambda}.
\end{equation}
In the PST approach this is extended to the manifestly covariant form
\begin{equation}
M_{\hat\mu\hat\nu}^{\rm cov.} = G_{\hat\mu\hat\nu} + i {G_{\hat\mu\hat\rho}
G_{\hat\nu \hat\lambda}\over\sqrt{-G (\partial a)^2}}
\tilde{H}_{\rm cov.}^{\hat\rho \hat\lambda}. \label{Mcov}
\end{equation}
The quantity
\begin{equation}
(\partial a)^2 = G^{\hat\mu\hat\nu} \partial_{\hat\mu} a \partial_{\hat\nu} a
\end{equation}
reduces to $G^{55}$ upon setting $\partial_{\hat\mu}a  = \delta_{\hat\mu}^5$,
and
\begin{equation}
\tilde{H}_{\rm cov.}^{\hat\rho\hat\lambda} \equiv {1\over 6} \epsilon^{\hat\rho
\hat\lambda \hat\mu \hat\nu \hat\sigma \hat\tau} H_{\hat\mu \hat\nu \hat\sigma}
\partial_{\hat\tau} a
\end{equation}
reduces to $\tilde{H}^{\rho\lambda}$.  Thus $M_{\hat\mu \hat\nu}^{\rm cov.}$
replaces $M_{\hat\mu\hat\nu}$ in $L_1$.  Furthermore, the expression
\begin{equation}
L' = - { 1\over 4(\partial a)^2} \tilde{H}_{\rm cov.}^{\hat\mu \hat\nu}
H_{\hat\mu\hat\nu\hat\rho} G^{\hat\rho\hat\lambda} \partial_{\hat\lambda} a,
\end{equation}
which transforms under general coordinate transformations as a scalar density,
reduces to $L_2 + L_3$ upon gauge fixing. It is interesting that $L_2$ and $L_3$ are
unified in this formulation.

Let us now describe the new gauge symmetries of ref.~\cite{pasti1}.  Since degrees of
freedom $a$ and $B_{\mu 5}$ have been added, corresponding gauge symmetries are
required.  One of them is
\begin{equation}
\delta B_{\hat\mu \hat\nu} = 2 \phi_{[\hat\mu} \partial_{\hat\nu]} a,
\end{equation}
where $\phi_{\hat\mu}$ are infinitesimal parameters, and the other fields do not
vary.  In terms of differential forms, this implies $\delta H = d\phi\wedge
da$.  $\tilde{H}_{\rm cov.}^{\hat\rho \hat\lambda}$ is invariant under this transformation,
since it corresponds to the dual of $H\wedge da$, but $da\wedge da = 0$.
Thus the covariant version of $L_1$ is invariant under this transformation.
The variation of $L'$, on the other hand, is a total derivative.

The second local symmetry involves an infinitesimal 
scalar parameter $\varphi$.  The transformation
rules are $\delta G_{\hat\mu\hat\nu} = 0$, $\delta a = \varphi$, and
\begin{equation}
\delta B_{\hat\mu\hat\nu} = {1\over (\partial a)^2} \varphi
H_{\hat\mu\hat\nu\hat\rho} G^{\hat\rho\hat\lambda} \partial_{\hat\lambda} a +
\varphi V_{\hat\mu\hat\nu},
\end{equation}
where the quantity $V_{\hat\mu\hat\nu}$ is to be determined.   
Rather than derive it from
scratch, let's see what is required to agree with the previous formulas after
gauge fixing.  In other words, we fix the gauge $\partial_{\hat\mu} a =
\delta_{\hat\mu}^5$ and $B_{\mu 5} = 0$, and figure out what the resulting
$\xi$ transformations are.  We need
\begin{equation}
\delta a = \varphi + \xi \partial_5 a = \varphi + \xi = 0,
\end{equation}
which tells us that $\varphi = - \xi$.  Then
\begin{eqnarray}
\delta_{\xi} B_{\mu\nu} &=& {1\over (\partial a)^2} \varphi H_{\mu\nu\hat\rho}
G^{\hat\rho\hat\lambda} \partial_{\hat\lambda} a + \varphi V_{\mu\nu} + \xi
H_{5\mu\nu}\nonumber \\
&=& - \xi \left({G^{\rho 5}\over G^{55}} H_{\mu\nu\rho} + V_{\mu\nu}\right) =
\xi (K_{\mu\nu}^{(\epsilon)} - V_{\mu\nu}).
\end{eqnarray}
Thus, comparing with eqs.~(\ref{Bvar}) and (\ref{Kform1}), we need the covariant definition
\begin{equation}
V_{\hat\mu\hat\nu} = - 2 {\partial L_1\over \partial
\tilde{H}_{\rm cov.}^{\hat\mu\hat\nu}}
\end{equation}
to achieve agreement with our previous results.

\medskip

\section{A New Heterotic String Action}

There are two main approaches to constructing the world-sheet action of the
heterotic string that have been used in the past~\cite{gross}.  
In one of them, the internal
torus is described in terms of bosonic coordinates. The fact that these
bosons are chiral ({\it i.e.}, the left-movers and right-movers behave differently)
is imposed through external constraints.  In the second approach these bosonic
coordinates are replaced by world-sheet fermions, which are Majorana--Weyl in
the 2d sense.  What will be most convenient for our purposes is a variant of
the first approach.  In this variant the coordinates of the Narain torus are still
represented by bosonic fields, but the chirality of these fields is achieved
through new gauge invariances rather than external constraints~\cite{cherkis}.

Consider the Narain compactified heterotic string in a
Minkowski space-time with $d = 10 - n$ dimensions~\cite{narain}.  
Let these coordinates be denoted
by $X^m$ with $m = 0,1, \ldots, d - 1 = 9 - n$.  To properly account for all
the degrees of freedom, the Narain torus should be described by $16 + 2n$
bosonic coordinates $Y^I, $ $I = 1,2, \ldots, 16 + 2n$.  These will be arranged to
describe $26 - d = 16 + n$ left-movers and $10 - d = n$ right-movers.  The
$Y^I$ are taken to be angular coordinates, with period $2\pi$, so that $Y^I
\sim Y^I + 2\pi$, and the conjugate momenta are integers.  The actual size and
shape of the torus is encoded in a matrix of moduli, denoted $M_{IJ}$, which
will be described below.

The $(16 + 2n)$-dimensional
lattice of allowed momenta should form an even self-dual lattice of signature
$(n, 16 + n)$.  Let us therefore introduce a matrix
\begin{equation}
\eta = \left(\begin{matrix}
I_n & 0\\
0 & - I_{16+n}\end{matrix}\right) ,
\end{equation}
where $I_n$ is the $n \times n$ unit matrix.  An
even self-dual lattice with this signature has a set of $16 + 2n$ basis
vectors $V_I$, and the symmetric matrix
\begin{equation}
L_{IJ} = V_I^a \eta_{ab} V_J^b
\end{equation}
characterizes the lattice.  A convenient specific choice is
\begin{equation}
L = \Lambda_8 \oplus \Lambda_8 \oplus \sigma \oplus \ldots \oplus \sigma,
\end{equation}
where $\Lambda_8$ is the negative of the $E_8$ Cartan matrix and $\sigma =
\left(\begin{matrix} 0 & 1\\ 1 & 0 \end{matrix}\right)$ appears $n$ times. 

The Narain moduli space is characterized, up to $T$ duality equivalences that
will be discussed below, by a symmetric matrix $M'_{ab} \in O (n, 16 +
n)$, which satisfies $M'\eta M' = \eta$. 
The fact that it is symmetric means that it actually parametrizes the
coset space $O(n, 16 + n)/O(n) \times O (16+n)$, which has $n(16 + n)$ real
dimensions.  To describe the $T$ duality equivalences, it is convenient to
change to the basis defined by the basis vectors of the self-dual lattice.
Accordingly, we define
\begin{equation}
M_{IJ} = V_I^a M'_{ab} V_J^b = (V^T M' V)_{IJ}.
\end{equation}
This matrix is also symmetric and satisfies
\begin{equation}
ML^{-1} M = L,
\end{equation}
from which it follows that $(L^{-1}M)^2 = 1$.  This allows us to define
projection operators
\begin{equation}
{\cal P}_\pm = {1\over 2} (1 \pm L^{-1} M).
\end{equation}
${\cal P}_+$ projects onto an $n$-dimensional subspace, which will correspond
to right-movers.  Similarly, ${\cal P}_-$ projects onto the
$(16+n)$-dimensional space of left-movers.
The theory we are seeking should be 
invariant under an infinite discrete group of $T$ duality
transformations, denoted 
$\Gamma_{n,16+n}$,\footnote{It is often called $O(n,16 + n; {\bf Z})$.} 
so that the actual moduli space is the standard Narain space
\begin{equation}
{\cal M}_{n, 16 + n} = \Gamma_{n, 16 + n}\backslash O(n, 16 + n) / O(n) \times O (16 +n).
\end{equation}

The desired equations of motion for the $Y$ 
coordinates are~\cite{cecotti,duff,tseytlin,maharana}
\begin{equation}
{\cal P}_- \partial_+ Y = 0 \quad { \rm and} \quad {\cal P}_+ \partial_- Y = 0,
\label{Yeqs}
\end{equation}
where $\xi^\pm = \xi^1 \pm \xi^0$, so that $\partial_\pm = {1\over 2}
(\partial_1 \pm \partial_0)$.  $\xi^0$ and $\xi^1$ are the world-sheet time and
space, respectively. The pair of equations in (\ref{Yeqs}) can be combined
in the form
\begin{equation}
M\partial_0 Y - L\partial_1 Y = 0.
\end{equation}
It is easy to write down a lagrangian that gives this equation~\cite{floreanini}:
\begin{equation}
{\cal L}_N = {1\over 2} (\partial_0 Y M \partial_0 Y - \partial_0 Y L \partial_1 Y).
\end{equation}
Two things are peculiar about this lagrangian.  First, it does not
have manifest Lorentz invariance.  
However, in ref.~\cite{schwarz} it
was shown that ${\cal L}_N$ has a global symmetry that can be interpreted as
describing a non-manifest Lorentz invariance. 
Second, it gives the equation of motion
\begin{equation}
\partial_0 [M\partial_0 Y - L\partial_1 Y] = 0,
\end{equation}
which has a second, unwanted, solution $Y^I = f^I (\xi^1)$.  The resolution of
the second problem is quite simple.  The transformation $\delta Y^I = f^I
(\xi^1)$ is a gauge symmetry of ${\cal L}_N$, and therefore $f^I (\xi^1)$ represents
unphysical gauge degrees of freedom.

The first problem, the noncovariance of ${\cal L}_N$, is more interesting.  
We will follow the PST approach~\cite{pasti2}, 
and extend ${\cal L}_N$ to a manifestly Lorentz
invariant action by introducing an auxiliary scalar field $a(\xi)$.  The
desired generalization of ${\cal L}_N$ is then
\begin{equation}
{\cal L}_{PST} = {1\over 2(\partial a)^2} (\tilde{Y} M \tilde{Y} - \tilde{Y} L\,
\partial Y \cdot \partial a),
\end{equation}
where
\begin{equation}
\tilde{Y}^I = \epsilon^{\alpha\beta} \partial_\alpha Y^I \partial_\beta a.
\label{Ytilde}
\end{equation}
Also, $(\partial a)^2$ and $ \partial Y \cdot \partial a$ are formed using the 2d
Lorentz metric, which  is diagonal with $\eta^{00} = - 1$ and $\eta^{11} = 1$.  

The theory given by ${\cal L}_{PST}$ has two gauge invariances.  The first is
\begin{eqnarray}
\delta Y &=& \varphi \left({1\over \partial_+ a} {\cal P}_- \partial_+ Y +
{1\over\partial_- a} {\cal P}_+ \partial_- Y\right),\nonumber \\
\delta a &=& \varphi,
\end{eqnarray}
where $\varphi (\xi^0, \xi^1)$ is an arbitrary infinitesimal scalar function.
If this gauge freedom is used to set $a = \xi^1$, then ${\cal L}_{PST}$ reduces to
${\cal L}_N$.  The second gauge invariance is
\begin{equation}
\delta Y^I = f^I (a), \quad \delta a = 0, \label{fifteen}
\end{equation}
where $f^I$ are arbitrary infinitesimal functions of one variable.  This is the
covariant version of the gauge symmetry of ${\cal L}_N$ that was used to argue that the
undesired solution of the equations of motion is pure gauge.

\medskip

\noindent{\it Reparametrization Invariant Action}

The formulas described above are not the whole story of the bosonic degrees of
freedom of the toroidally compactified heterotic string, because they lack the
Virasoro constraint conditions.  The standard way to remedy this situation is
to include an auxiliary world-sheet metric field 
$g_{\alpha\beta}(\xi)$, so that the world-sheet
Lorentz invariance is replaced by world-sheet general coordinate invariance.
Since we now want to include the coordinates $X^m$ describing the uncompactified
dimensions, as well, let us also introduce an induced world-sheet metric
\begin{equation}
G_{\alpha\beta} = g_{mn}(X) \partial_{\alpha}X^m\partial_{\beta} X^n,
\end{equation}
where $g_{mn}(X)$ is the string frame target-space metric in $d$ dimensions.
It is related to the canonically normalized metric by a factor of the form
${\rm exp} (\alpha\phi)$, where $\phi$ is the dilaton and $\alpha$ is a numerical
constant, which can be computed by requiring that the target-space lagrangian is
proportional ${\rm exp} (-2\phi)$.  We will mostly be
interested in taking $\phi$ to be a constant and $g_{mn}$ to be proportional
to the flat Minkowski metric. Then the heterotic string coupling constant
is $\lambda_H = {\rm exp}\, \phi$, and the desired world sheet lagrangian is
\begin{equation}
{\cal L}_g = - {1\over 2} \sqrt{-g} g^{\alpha\beta} G_{\alpha\beta}
+ {\tilde{Y} M \tilde{Y}\over 2\sqrt{-g} (\partial a)^2} - {\tilde{Y} L\,
\partial Y \cdot \partial a\over 2(\partial a)^2} . \label{Laux}
\end{equation}
Now, of course, $(\partial a)^2 = g^{\alpha\beta} \partial_\alpha a
\partial_\beta a$ and $ \partial Y \cdot \partial a = g^{\alpha\beta}
\partial_\alpha Y \partial_\beta a$.  The placement of the $\sqrt{-g}$ factors
reflects the fact that $\tilde{Y}/\sqrt{-g}$ transforms as a scalar. 

There are a few points to be made about ${\cal L}_g$.  First of all, the PST gauge
symmetries continue to hold, so it describes the correct degrees of freedom.
Second, just as for more conventional string actions, it has Weyl invariance:
$g_{\alpha\beta} \rightarrow \lambda g_{\alpha\beta}$ is a local symmetry.
This ensures that the stress tensor
\begin{equation}
T_{\alpha\beta} = - {2\over\sqrt{-g}} {\delta S_g\over\delta g^{\alpha\beta}},
\end{equation}
is traceless $(g^{\alpha\beta} T_{\alpha\beta} = 0)$.
Using the general coordinate invariance to choose $g_{\alpha\beta}$ conformally
flat, and using the PST gauge invariance to set $a = \xi^1$, the $Y$ equations
of motion reduce to those described in the previous subsection.  In addition,
one obtains the classical Virasoro constraints $T_{++} = T_{--} = 0$.  

The lagrangian ${\cal L}_g$ is written with an auxiliary world-volume metric, which is
called the Howe--Tucker or Polyakov formulation.  This is the most convenient
description for many purposes.  However, for the purpose of comparing to
expressions derived from the M5-brane later in this paper, it will be useful to
also know the version of the lagrangian in which the auxiliary metric is
eliminated --- the Nambu--Goto formulation.
Note that ${\cal L}_g$ only involves the metric components in the combination
$\sqrt{-g} g^{\alpha\beta}$, which has two independent components.
It is a straightforward matter to solve their equations of motion and 
eliminate them from the action. This leaves
the final form for the bosonic part of the heterotic string in
$10 - n$ dimensions
\begin{equation}
{\cal L} = - \sqrt{-G}\sqrt{1 + {\tilde{Y} M \tilde{Y}\over G(\partial a)^2} +
\left({\tilde{Y} L \tilde{Y}\over 2G(\partial a)^2}\right)^2 }- {\tilde{Y}
L\partial Y \cdot \partial a\over 2(\partial a)^2}, \label{hetfinal}
\end{equation}
where $G = \det G_{\alpha\beta}$, and now
\begin{equation}
(\partial a)^2 = G^{\alpha\beta} \partial_\alpha a \partial_\beta a, \quad \partial Y
\cdot \partial a = G^{\alpha\beta} \partial_\alpha Y \partial_\beta a.
\end{equation}

\medskip
\section{Wrapping the M-Theory Five-Brane on K3}

Let us now consider double dimensional reduction of the M5-brane 
on K3.\footnote{See ref.~\cite{aspinwall} for a review of the mathematics of K3
and some of its appearances in string theory dualities.} 
This is supposed to give the heterotic string in seven dimensions~\cite{witten,harvey,townsend}.
Our starting point is the bosonic part of the M5-brane action~\cite{perry}
in the general coordinate invariant PST formulation. 
Since the other 11d fields are still assumed to vanish, 
$g_{MN}(X)$ must be Ricci flat. We will take it to be a product of a Ricci-flat
K3 and a flat 7d Minkowski space-time.

Since the M5-brane is taken to wrap the spatial K3, 
the diffeomorphism invariances of
the M5-brane action in these dimensions can be used to equate 
the four world-volume  coordinates  that describe the K3 
with the four target-space coordinates that describe the K3. In other words,
we set $\sigma^{\mu} = ( \xi^{\alpha}, x^i)$ and $X^M = (X^m, x^i)$.
Note that Latin indices $i,j,k$ are used for the K3 dimensions $(x^i)$ 
and early Greek letters for the directions $(\xi^\alpha)$,
which are the world-sheet coordinates of the resulting string action.
This wrapping by identification of coordinates, together with the extraction of the
K3 zero modes, is what is meant by double dimensional reduction. 
With these choices, the 6d metric can be decomposed into blocks
\begin{equation}
\label{metricdecomposition}
\left(G_{\mu\nu}\right)=\left( \begin{array}{cc} 
                                    \tilde G_{\alpha\beta} & 0 \\ 
                                      0   &  h_{ij}    
                                          \end{array} \right) ,
\end{equation}
with $h_{ij}$ and $\tilde G_{\alpha\beta}$ being the K3 metric and the induced
metric on the string world-sheet, respectively. The purpose of the tilde is to
emphasize that $\tilde G_{\alpha\beta} = \tilde g_{mn} \partial_{\alpha} X^m
\partial_{\beta} X^n$, where $\tilde g_{mn}$ is the 7d part of the
canonical 11d metric. It differs from the metric introduced earlier by a scale factor,
which will be determined below. It is
convenient to take the PST scalar field $a$
to depend on the $\xi^{\alpha}$ coordinates only. This amounts to partially
fixing a gauge choice for the PST gauge invariance.

The two-form field $B$ has the following contributions from K3 zero modes:
\begin{equation}
\label{2formdecomposition}
B_{ij}= \sum_{I=1}^{22} Y^I(\xi) b_{Iij}(x), \;\; B_{\alpha i}=0, 
\;\; B_{\alpha\beta}=c_{\alpha\beta}(\xi),
\end{equation}
where $b_{Iij}$ are the 22 harmonic representatives of H${}^2$(K3, {\bf Z}), the
integral second cohomology classes of K3. Any other terms are either massive or can be
removed by gauge transformations.
The nonzero components of $H_{\mu\nu\rho}$ and $\Htilde^{\mu\nu}$ are
\begin{equation}
H_{\alpha i j}=\sum_{I=1}^{22} \partial_\alpha Y^I b_{Iij} \label{Hform1}
\end{equation}
\begin{equation}
\Htilde^{i j}=\sum_{I=1}^{22}\Ytilde^I \frac{1}{2}\ep^{i j k l}b_{Ik l}=
\sum_{I=1}^{22}\sqrt{h} \Ytilde^I (\ast b_I)^{i j}, \label{Hform2}
\end{equation}
where $\Ytilde^I=\ep^{\alpha\beta}\partial_\alpha Y^I \partial_\beta a$ as in 
eq.~(\ref{Ytilde}). Note that $c_{\alpha\beta}$ does not contribute.

Now we can compute the string action that arises from double dimensional
reduction by substituting the decompositions (\ref{Hform1}) and (\ref{Hform2})
into the five-brane Lagrangian. 
To make the connection with the heterotic string action of the previous section,
we make the identifications 
\begin{equation}
 L_{IJ}=\int_{K3} b_I\wedge b_J,
\end{equation}
\begin{equation}
M_{IJ}=\int_{K3} b_I\wedge \ast b_J.
\end{equation}
Note that $\ast b_I = b_J (L^{-1} M)^J{}_I$, and therefore $(L^{-1}M)^2 =1$,
as in sect. 2.
Note also that $b_I\wedge b_J$ and $b_I\wedge \ast b_J$ are closed four-forms, and
therefore they are cohomologous to the unique harmonic four-form of the K3, 
which is the volume form $\omega$. It follows that
\begin{equation} \label{bwedgeb}
b_I\wedge b_J=\ast b_I\wedge \ast b_J=\frac{L_{IJ}}{\cal V}\omega +dT_{IJ},\quad  
b_I\wedge \ast b_J=\frac{M_{IJ}}{\cal V}\omega +dU_{IJ},
\end{equation}
where ${\cal V}= \int_{K3} \omega$ is the volume of the K3
and $U_{IJ} =  T_{IK}(L^{-1} M)^K{}_J$. The exact terms are absent when either
two-form is self-dual, but there is no apparent reason why they should vanish when
both of them are anti-self-dual.
If we nevertheless ignore the exact pieces in these formulas,
substitute into  the Lagrangian,
and integrate over the K3, we obtain
\begin{equation}
\label{5brane on K3}
{\cal L}_1 = -{\cal V} \sqrt{-\tilde G}\sqrt{1+
\frac{\Ytilde^I M_{IJ} \Ytilde^J}{\tilde G(\partial a)^2 {\cal V}}
+\frac{1}{4}\left(\frac{\Ytilde^I L_{IJ}\Ytilde^J}
{\tilde G(\partial a)^2{\cal V}} \right)^2}
-\frac{\Ytilde^I L_{IJ} \partial_\alpha Y^J \tilde G^{\alpha\beta}\partial_\beta a}
{2(\partial a)^2}.
\end{equation}
This is precisely the heterotic string lagrangian (for $n=3$) presented in 
eq.~(\ref{hetfinal}) of the previous section provided that the 7d metric $g_{mn}$
in the string frame is related to the metric $\tilde g_{mn}$ derived from 11d by 
\begin{equation}
g_{mn} = {\cal V} \tilde g_{mn}
\end{equation}
so that $G_{\alpha\beta} = {\cal V}
\tilde G_{\alpha\beta}$. This is the same scaling rule found by a different 
argument in ref.~\cite{witten}.  Then, following ref.~\cite{witten},
the Einstein term in the 7d lagrangian
is proportional to ${\cal V} \sqrt{- \tilde g} R(\tilde g) = {\cal V}^{-3/2}
\sqrt{- g} R(g)$, from which
one infers that ${\cal V} \sim \lambda_H^{4/3}$.

To complete the argument we must still explain why terms that have been dropped
make negligible contributions. It is not at all obvious that 
the exact pieces in eq.~(\ref{bwedgeb}) can be neglected, but it is what is required
to obtain the desired answer. The other class of terms that have been dropped are
the Kaluza--Klein excitations of the five-brane on the K3. By simple dimensional
analysis, one can show that in the heterotic string metric these contributions 
to the mass-squared of excitations are of order $\lambda_H^{-2}$. Therefore they
represent non-perturbative corrections from the heterotic viewpoint. Since our
purpose is only to reproduce the perturbative heterotic theory, they can be dropped.
Another class of contributions, which should not be dropped, correspond to
simultaneously wrapping the M2-brane around a 2-cycle of the K3. These wrappings
introduce charges for the 22 U(1)'s, according to how many times each cycle is wrapped.
The contribution to the mass-squared of excitations depends on the shape of the K3,
of course, but in the heterotic metric it is independent of its volume and hence of the
heterotic string coupling constant.

\noindent{\it Acknowledgment}

I am grateful to M. Aganagic, S. Cherkis, 
J. Park, M. Perry, and C. Popescu for collaborating on portions of this work.
This work is 
supported in part by the U.S. Dept. of Energy under Grant No.
DE-FG03-92-ER40701.

\medskip

\section*{Appendix :  Solution of eq.~(\ref{nlpde})}

The differential
equation
\begin{equation}
f_1^2 + y_1 f_1 f_2 + \left({1\over 2} y_1^2 - y_2\right) f_2^2 = 1
 \label{diffeqn}
\end{equation}
can be made to look much simpler by the
change of variables
\begin{eqnarray}
y_1 &=& - (u_+ + u_-)\nonumber \\
y_2 &=& {1\over 2} (u_+^2 + u_-^2).
\end{eqnarray}
Denoting the resulting function by the same symbol, $f(u_+, u_-)$, and
derivatives by $f_\pm \equiv {\partial f\over \partial u_\pm}$, one has
\begin{eqnarray}
f_1 &=& {u_- f_+ - u_+ f_-\over u_+ - u_-}\nonumber \\
f_2 &=& {f_+ - f_-\over u_+ - u_-}.
\end{eqnarray}
Substituting these in eq. (\ref{diffeqn})  then gives the remarkably simple
differential equation
\begin{equation}
f_+ f_- = 1.
\end{equation}
Essentially the same equation was discovered in Ref. \cite{gibbons}
 as the condition for
electric-magnetic duality symmetry of a 4d $U(1)$ gauge theory.  Perhaps,
in retrospect, this
is not too surprising.

Fortunately, the general solution of the equation $f_+ f_- =1$
 is given in Courant
and Hilbert \cite{courant}.   It is given parametrically in terms of
 an arbitrary function
$v(t)$:
\begin{eqnarray}
f &=& {2u_+\over \dot v(t)} + v(t)\nonumber \\
u_- &=& {u_+\over (\dot v(t))^2} + t,
\end{eqnarray} 
where the dot means that the derivative of the function is taken
 with respect to its argument.
In principle, the second equation determines $t$ in terms of $u_+$ and $u_-$,
which can then be substituted into the first one to give $f$ in terms of $u_+$
and $u_-$.  The proof is simple, so we show it.  Taking differentials,
\begin{eqnarray}
df &=& {2\over\dot v} du_+ + \left(\dot v - {2\ddot v\over \dot v^2}u_+\right)
dt\nonumber \\
du_- &=& {1\over (\dot v)^2} du_+ + \left( 1 - {2\ddot v\over \dot
v^3}u_+\right) dt.
\end{eqnarray}
Eliminating $dt$ leaves
\begin{equation}
df = {1\over \dot v} du_+ + \dot v du_-,
\end{equation}
which implies that $f_+ = 1/\dot v$ and $f_- = \dot v$, so that $f_+ f_- = 1$.

This is not the whole story, since  $f(y_1, y_2)$ is required to be analytic at the
origin.  This implies that
\begin{equation} \label{symcond}
f (u_+, u_-) = f(u_-, u_+).
\end{equation}
Letting
$\varphi (t) = \dot v (t)$ and $\psi(t) = - t \varphi^2 (t)$ eq. (\ref{symcond})
implies that
\begin{equation}
\psi (\psi(t)) = t. \label{psieqn}
\end{equation}
In words, the function is the same as the inverse function.  

Large classes of solutions of (\ref{psieqn}) are obtained as follows. Pick
a symmetric function $F(s,t) = F(t,s)$ and determine $\psi(t)$ by 
\[ F(\psi, t) =0. \]
For example, the simplest non-trivial choice is
\[ F(s,t) = s + t + \alpha s t ,\]
which gives
\begin{equation}
\psi (t) = {- t\over 1 + \alpha t} .
\end{equation}
Choosing the normalization $\alpha =1$ gives the solution cited in the text as the
one that is relevant to the M theory fivebrane. Other solutions may be of interest for
other purposes.

\end{document}